# Incoherent Effect of Fe and Ni Substitutions in the Ferromagnetic-Insulator $La_{0.6}Bi_{0.4}MnO_{3+\delta}$

**Asish K. Kundu[1]\*, Md. Motin Seikh[2], Akhilesh Srivastava[1], S. Mahajan[3], R. Chatterjee[3]**

**V. Pralong[4] and B. Raveau[4]**

[1]*Indian Institute of Information Technology Design & Manufacturing, Dumna Airport Road, Jabalpur – 482005, India*

[2]*Department of Chemistry, Visva-Bharati University, Santiniketan–731235, India*

[3]*Department of Physics, Indian Institute of Technology, Hauz Khas, New Delhi-11016, India*

[4]*CRISMAT Laboratory, ENSICAEN UMR6508, 6 Bd Maréchal Juin, Cedex 4, Caen-14050, France*

**Abstract**

A comparative study of the effect of Fe and Ni doping on the bismuth based perovskite $La_{0.6}Bi_{0.4}MnO_{3.1}$, a projected spintronics magnetic semiconductor has been carried out. The doped systems show an expressive change in magnetic ordering temperature. However, the shifts in ferromagnetic transition ($T_C$) of these doped phases are in opposite direction with respect to the parent phase $T_C$ of 115 K. The Ni-doped phase shows an increase in $T_C$ ~200 K, whereas the Fe-doped phase exhibits a downward shift to $T_C$~95 K. Moreover, the Fe-doped is hard-type whereas the Ni-doped compound is soft-type ferromagnet. It is observed that the materials are semiconducting in the ferromagnetic phase with activation energies of 77 & 82 meV for Fe & Ni-doped phases respectively. In the presence of external magnetic field of 7 Tesla, they exhibit minor changes in the resistivity behaviours and the maximum isothermal magnetoresistance is around -20 % at 125 K for the Ni-phase. The results are explained on the basis of electronic phase separation and competing ferromagnetic and antiferromagnetic interactions between the various mixed valence cations.

\*Corresponding author: asish.k@gmail.com/asish.kundu@iiitdmj.ac.in

**I. INTRODUCTION**

In the last few decades there has been an increasing interest in the understanding of the basic physics/chemistry of spintronics materials.[1,2] Despite the numerous investigations on spintronics materials in the past few years very few perovskite-manganites have been known as ferromagnetic-insulators (FMI).[1-6] In this respect, the perovskites $La_{1-x}Bi_xMnO_{3+\delta}$ studied for their FMI properties[7,8] and recently discovered multiferroics $La_{0.1}Bi_{0.9}MnO_3$ (Ref. 9) and $La_{0.2}Bi_{0.8}MnO_3$ (Ref. 10) are of great interest. These La-substituted $BiMnO_3$ phases show multiferroic properties also at low temperature (T ≈ 100 K), similar to the parent compound. It is noticed that for Bi rich ($x \geq 0.6$), high-pressure/temperature synthesis is required and the multiferroicity is reported for $x$ = 0.8 and 0.9 phases only in the epitaxial thin films.[7-10] $La_{0.1}Bi_{0.9}MnO_3$ ($La_{0.2}Bi_{0.8}MnO_3$) exhibits ferroelectricity at around 300 K (150 K) and ferromagnetism around ~95K (90 K), but for the bulk phase there is no such report.[9,10]

The interplay between magnetic, electronic and structural properties gives rise to fascinating complex phenomena and therefore the basic physics of the materials is rich. In recent years we have been exploring the kind of Bi-based manganite perovskites, which show simultaneously ferromagnetic (FM) and insulating properties.[11,12] Actually, the ferromagnetic-semiconductors/insulators are novel singular materials that could exhibit simultaneously electric and magnetic ordering.[6] Such materials have practical application in spintronics and magneto-dielectric based devices.[5,6] Also, the recent investigations on multiferroic and/or spin filtering in this type of thin films have enhanced the possibility of device applications.[9,10]

In this paper, we have studied the magneto-transport properties of the Bi-based manganite where $Mn^{3+}$ is replaced by Fe/Ni-ions. Bearing in mind that the substitution of bismuth for lanthanum in $LaMnO_{3+\delta}$ increases the insulating properties but tends to decrease $T_C$,[7, 8] we have selected the $La_{0.6}Bi_{0.4}Mn_{0.6}(Fe/Ni)_{0.4}O_{3+\delta}$ series and varied the Fe/Ni level so that we can achieve ferromagnetism with insulating properties. Interestingly, it is observed that the FM $T_C$ is increased from 115 K (for $La_{0.6}Bi_{0.4}MnO_{3.10}$) to 200 K for $La_{0.6}Bi_{0.4}Mn_{0.6}Ni_{0.4}O_{2.94}$ and decreases to 95 K for $La_{0.6}Bi_{0.4}Mn_{0.6}Fe_{0.4}O_{3.1}$.

## II. EXPERIMENTAL PROCEDURE

Polycrystalline samples were prepared by conventional sol-gel method.[11] Finally the powder samples were pressed into rectangular bars and sintered, at 1173K and 1223K for $La_{0.6}Bi_{0.4}Mn_{0.6}Ni_{0.4}O_{2.94}$ and $La_{0.6}Bi_{0.4}Mn_{0.6}Fe_{0.4}O_{3.1}$ respectively, in platinum crucible for 24h to obtain the single phase compounds. Phase purity was checked by X-ray powder diffraction (XRPD) measurements and oxygen stoichiometry was determined by chemical titration for all samples as described in Ref. 11. The magnetization, resistivity and magnetoresistance measurements were carried out with a Quantum Design physical properties and magnetic properties measurements systems (PPMS; MPMS). The details of the magnetization and electrical resistivity measurements procedure is mentioned in our previous work.[11, 12]

## III. RESULTS AND DISCUSSION

Figure 1 shows the XRPD patterns for both samples $La_{0.6}Bi_{0.4}Mn_{0.6}(Fe/Ni)_{0.4}O_{3+\delta}$. Structural parameters in the orthorhombic space group *Pnma* were refined by the Rietveld method using the program RIETAN 2000.[13] Good reliability factors similar to those of our previous studies were obtained[11, 12] and are presented in table 1. Insets of Fig. 1 depict the scanning electron microscope images of the polycrystalline pellets. It is observed that the average particle size for Fe-doped sample is 1-5μm whereas clusters of phases appear in the case of Ni-doped sample. Hence the surface morphology for both samples has a clear distinct feature, which may influence the physical properties of the systems.

The temperature dependence of zero field cooled (ZFC) and field cooled (FC) magnetization data, M(T), of $La_{0.6}Bi_{0.4}Mn_{0.6}(Fe/Ni)_{0.4}O_{3+\delta}$ in an applied field of 0.01T are shown in Fig. 2. The $La_{0.6}Bi_{0.4}Mn_{0.6}Fe_{0.4}O_{3.1}$ compound shows a paramagnetic (PM) to ferromagnetic (FM) transition, $T_C$, around 95 K [Fig. 2(a)] similar to the parent phase $La_{0.6}Bi_{0.4}MnO_{3.1}$.[7, 8] Additionally with decreasing the temperature, the M(T) data show a FM to antiferromagnetic (AFM)-like transition around 60 K, where the ZFC magnetization decreases rapidly to a lower constant value. Moreover, the FC curve exhibits an upturn around 50 K with a large thermomagnetic irreversibility between them. On the other hand, the Ni-substitution results in a significant increase of the FM $T_C$ to 200 K. The ZFC and FC magnetization data depict large divergence for both samples, for a lower applied field (≤ 100 Oe), akin to the reported compound.[11, 12] This suggests that, the magnetic interactions in Fe/Ni-doped samples are due to short-range type ferromagnetism. To further understand the nature of different types of magnetic interactions we have carried out low temperature magnetic measurements in details, which will be discussed later. In Fig. 2, we have also shown the ZFC and FC magnetization data of $La_{0.6}Bi_{0.4}Mn_{0.6}(Fe/Ni)_{0.4}O_3$, studied in higher applied fields (0.1 and 0.5 T). For this Ni-phase the irreversible temperature ($T_{irr}$) obtained from the ZFC and FC magnetization curves is $T_{irr}$ ~179 K and two broad magnetic transitions appear at 90 and 170 K (H = 0.01 T), which are also present in higher fields but the transitions are rather faint [Fig. 2(b)]. With increasing the field value, $T_{irr}$ shifts towards lower temperature and the curves merge for sufficiently higher field (H = 0.1T). In contrast, for the $La_{0.6}Bi_{0.4}Mn_{0.6}Fe_{0.4}O_{3.1}$ sample, with a higher field of 0.1 T, both the ZFC and FC curves merge above the AFM-like transition ($T_{irr} \sim 60$ K) and the ZFC cusp remains prominent [Figs. 2(a)], similar to that of a spin or cluster glass type material as reported in the literature.[14-16]

In the paramagnetic (PM) region ($T > T_C$) the magnetization curve is well fitted to the Curie-Weiss law. The temperature dependence inverse magnetic susceptibility data in the temperature range of 10-400 K (Fig. 3), follows Curie-Weisss behaviour above $T_C$ and from the linear fitting we have calculated the PM Weiss temperature ($\theta_p$) and the effective magnetic moment ($\mu_{eff}$). The $\theta_p$ values for $La_{0.6}Bi_{0.4}Mn_{0.6}Fe_{0.4}O_{3.1}$ and $La_{0.6}Bi_{0.4}Mn_{0.6}Ni_{0.4}O_{2.94}$ compounds are 34 K and 200 K, whereas the $\mu_{eff}$ values are 3.17 $\mu_B$/f.u and 2.85 $\mu_B$/f.u. respectively. The obtained $\theta_p$ values are positive, which confirms the existence of FM interactions in the high temperature region as well.

The low temperature magnetic phase has been investigated in details to characterize the nature of FM interactions between the different magnetic ions such as $Mn^{3+}$, $Mn^{4+}$, $Fe^{3+}$ or $Ni^{2+}$. The isothermal magnetic hysteresis loops, M(H), have been studied at different temperatures (Fig. 4). For the $La_{0.6}Bi_{0.4}Mn_{0.6}Ni_{0.4}O_{2.94}$ sample, one observes [Fig. 4(b)] a very small hysteresis loop at 10 K, with a remanent magnetization ($M_r$) value of ~0.31 $\mu_B$/f.u. and a coercive field ($H_C$) of ~ 0.0085 T. These values are almost similar to the parent $La_{0.6}Bi_{0.4}MnO_{3.1}$ phase.[12] At low temperature, the relatively smaller value of $H_C$ indicates a soft FM type behavior and above $T_C$ the M(H) behaviour is linear, corresponding to a PM state. In contrast, the hysteresis loops for $La_{0.6}Bi_{0.4}Mn_{0.6}Fe_{0.4}O_{3.1}$ are relatively large with a higher coercive field of 0.25 T at 10 K and the coercive value decreases gradually with increasing temperature [Fig. 4(a)]. Moreover we have noticed S-shape hysteresis curves for this compound below $T_C$ akin to glassy ferromagnetic system.[14-16] The highest values of magnetic moment is ~ 0.7 $\mu_B$/f.u. for $La_{0.6}Bi_{0.4}Mn_{0.6}Fe_{0.4}O_{3.1}$ and of 2.1 $\mu_B$/f.u. for $La_{0.6}Bi_{0.4}Mn_{0.6}Ni_{0.4}O_{2.94}$ are much smaller than the spin-only values corresponding to magnetic Fe(Ni)/Mn-ions. This can be explained by the fact that the FM and AFM components coexist at low temperature. Indeed the data show the superposition of two types of contributions below the FM transitions. The FM component is characterized by a hysteresis loop and a finite value of coercive field and the AFM component is characterized by unsaturated value of the magnetization. The latter increases almost linearly with increasing applied field. The observed M(H) behavior in these compounds is due to the presence of different magnetic interactions. Therefore, at low temperatures there must be a subtle balance between the FM and AFM interactions or in other words the system is electronically phase separated into FM and AFM clusters. Hence, there will always be a competition between these two interactions to dominate one over another giving rise to a tendency of glassy FM state in the material.[11, 15, 16] The fact that the magnetic moment (2.1 $\mu_B$/f.u.) is higher for $La^{III}_{0.6}Bi^{III}_{0.4}Mn^{III}_{0.32}Mn^{IV}_{0.28}Ni^{II}_{0.4}O^{II}_{2.94}$ may be due to the fact that this oxide exhibits a higher FM contribution due to the dominant $Mn^{4+}$-O-$Ni^{2+}$ FM interactions, combined with FM $Mn^{3+}$-O-$Mn^{4+}$ interactions. In contrast, the lower magnetic moment of $La^{III}_{0.6}Bi^{III}_{0.4}Mn^{III}_{0.4}Mn^{IV}_{0.2}Fe^{III}_{0.4}O^{II}_{3.1}$ (0.7 $\mu_B$/f.u.) suggests that the main FM contribution is due to $Mn^{3+}$-O-$Mn^{4+}$ interactions only, the magnetic $Mn^{3+}$-O-$Fe^{3+}$ interactions being most probably weak. In any case, an interesting feature at low temperature is the unsaturated behavior of the M(H) curves, even at higher fields for all samples, which is a characteristic feature of glassy-ferromagnets.[11, 15]

In order to further characterize the low temperature magnetic phase of the $La_{0.6}Bi_{0.4}Mn_{0.6}(Fe/Ni)_{0.4}O_{3+\delta}$ systems, we have carried out frequency-dependent magnetic measurement which is a very efficient way to investigate the magnetic glassy behavior. Fig. 5 (a-d) shows the temperature dependent in-phase $\chi'(T)$ and out-of-phase $\chi''(T)$ components of the ac-susceptibility for $La_{0.6}Bi_{0.4}Mn_{0.6}Ni_{0.4}O_{2.94}$ and $La_{0.6}Bi_{0.4}Mn_{0.6}Fe_{0.4}O_{3.1}$ measured at four different frequencies. The in-phase $\chi'(T)$ data exhibit similar features to the low field ZFC magnetization data. For $La_{0.6}Bi_{0.4}Mn_{0.6}Ni_{0.4}O_{2.94}$ the abrupt increase of $\chi'(T)$ at ~190 K characteristic of a FM ordering, is almost frequency-independent [Fig. 5(a)]. This behavior suggests that below $T_C$ the FM state originates from intra-cluster ferromagnetism i.e. from cluster glass type behavior rather than typical long-range ferromagnetism. Nevertheless, $La_{0.6}Bi_{0.4}Mn_{0.6}Fe_{0.4}O_{3.1}$ system depicts [Fig. 5(c, d)] a clear frequency-dependent peak below $T_C$ (~70 K) which shifts towards higher temperature with increasing frequency similar to $La_{0.6}Bi_{0.4}Mn_{0.6}Co_{0.4}O_{3.01}$ phase.[12] From this available data we can conjecture the possible existence of spin-glass behavior at low temperatures for the Fe doped samples though it require further investigations to confirm this glassy magnetic phase.[15]

Furthermore, we have carried out temperature dependent electrical resistivity measurements $\rho(T)$ for both samples. Fig. 6 shows the $\rho(T)$ in presence and absence of the external magnetic field of 7 T. In the zero applied field condition, the resistivity increases rapidly for both of them and they are insulating in nature throughout the measured temperature range (80 K ≤ T ≤ 400 K). The resistivity value is very high at low temperature, crossing instrument limitations. We have studied $\rho(T)$ in an applied field of 7 T and observed that the insulating behavior is robust for both the compounds similar to reported Bi-based perovskites.[12] However, at low temperature the $\rho(T)$ value slightly diverge in the presence of magnetic field for both the compounds and the effect is significantly weak for $La_{0.6}Bi_{0.4}Mn_{0.6}Fe_{0.4}O_{3.1}$, although the effect is not clear from Fig. 6. To further characterize this magnetotransport correlation we have investigated the isothermal MR measurements at different temperatures. It is worthy to mention that both the samples are FMI below room temperature and consequently, though the MR effect is small in magnitude, will be discussed in the next section. Thus, at low temperatures a strong correlation between magnetic and electric properties is observed in these systems.

Fig. 7 shows the isothermal magnetoresistance (MR) behavior of $La_{0.6}Bi_{0.4}Mn_{0.6}Fe_{0.4}O_{3.1}$ and $La_{0.6}Bi_{0.4}Mn_{0.6}Ni_{0.4}O_{2.94}$ at different temperature regions. The MR value is calculated as MR (%) = $[\{\rho(7)-\rho(0)\}/\rho(0)]\times100$, where $\rho(0)$ is the samples resistivity at 0 T and $\rho(7)$ is the resistivity under an applied field of ±7 Tesla. It is observed that the obtained MR values are negative and consistent with the values reported for Bi-based perovskite.[8, 12] The maximum negative MR values (at 125 K) are around 2.5 % and 20 % in an applied field of ±7 T for $La_{0.6}Bi_{0.4}Mn_{0.6}Fe_{0.4}O_{3.1}$ and $La_{0.6}Bi_{0.4}Mn_{0.6}Ni_{0.4}O_{2.94}$ respectively (the data was beyond the instrumental limit for T ≤ 120 K). It is observed that for $La_{0.6}Bi_{0.4}Mn_{0.6}Ni_{0.4}O_{2.94}$ the MR values are irreversible in nature and the occurrence of anisotropic MR behavior at low temperatures similar to those in M(H) studies suggests a strongly correlated nature of field-induced magnetic and electronic behavior. Nevertheless, the obtained MR values are less than the expected values below the FM $T_C$.

It is well known that in compounds with competing magnetic orders, a magnetic field favoring one kind of order in the spins also causes a large negative MR. For the present systems with competing FM and AFM interactions, field induced negative magneto-transport interactions are thus expected. The obtained negative MR for all compounds can be well understood in the scenario of suppression of the electron scattering below the FM $T_C$ in the presence of an external field. The charge transport in this system turns out to be very sensitive to the FM ordering, and the magnetic fields readily induce a negative MR below $T_C$. Hence, all compounds show FM and insulating behavior (even at high magnetic field) with electronic phase separation at low temperature (FM and AFM clusters).

In order to throw light on the transport mechanisms operating in the present compounds, we have further analyzed the low temperature behavior. Different models[12, 18] have been proposed to describe the temperature dependent charge transport behavior for oxide systems defined by $\log \rho \propto T^{-1/n}$ where n = 1, 2 or 4. These models are compared to describe the zero field $\rho(T)$ behavior in our compounds. Fig. 8(a) shows that the TA model, $\log \rho \propto 1/T$, describes well the resistivity behaviour of our samples above the FM transition $T_C$ with the activation energy, $E_a$, of 77 and 82 meV for $La_{0.6}Bi_{0.4}Mn_{0.6}Fe_{0.4}O_{3.1}$ and $La_{0.6}Bi_{0.4}Mn_{0.6}Ni_{0.4}O_{2.94}$ respectively. This is consistent with the value reported earlier for Bi-based manganites.[12] This suggests that increasing the doping concentration at Mn-site the energy band gap increases gradually. It is clearly noticed from Fig. 8(b) that, the SPH model, $\log \rho \propto T^{-1/2}$, aptly describes the resistivity behavior for the $La_{0.6}Bi_{0.4}Mn_{0.6}Ni_{0.4}O_{2.94}$ compound in the temperature range of 100 K ≤ T ≤ 400 K, among the three classes of model generally applied to the perovskite materials.[12] The SPH type conductivity is usually observed when the coulomb interaction starts to play a key role in carriers hopping. The SPH or VRH models do not fit to the experimental data at high temperatures for the $La_{0.6}Bi_{0.4}Mn_{0.6}Fe_{0.4}O_{3.1}$ compound in contrast to the TA model. It must be emphasized that for Fe/Ni-doped samples, the resistivity evolution follows the VRH model only in the low temperature region (T ≤ 220 K) as shown in the inset of Fig. 8(b). This type of hopping conduction is typical to the phase separated systems (FM and AFM) where the charge carriers move by hopping between two localized electronic states. Then the conductivity results from hopping motion of the charge carriers through such localized states.[12, 17]

## IV. CONCLUSIONS

The magnetic and electrical properties of the bismuth based perovskite $La_{0.6}Bi_{0.4}Mn_{0.6}M_{0.4}O_{3+\delta}$ (M = Fe and Ni) have been investigated in details. The effect of Fe and Ni doping is completely reverse for magnetic properties, the FM $T_C$ increases in the case of Ni and decreases for Fe substitution, the saturation moments become lower than that of the parent phase. The complex interplay between the mixed valency of the doped phases leads to competing FM and AFM interactions, which demonstrates the signature of electronic phase separation where the likely glassy ferromagnetic state appears. We have been successful to uplift the FM transition of the Ni-doped phase retaining the insulating behaviour. The substitution of manganese by Fe and Ni gradually increases the band gap, yet the MR effects are lower in the doped phases. Finally, the present investigation demonstrates the strongly correlated feature of the magnetic and electronic properties in these materials.

## Acknowledgements

AKK gratefully acknowledge Prof. A. Ojha for providing the faculty research grant.

**Table. 1.** Lattice parameters, magnetic and electrical properties of $La_{0.6}Bi_{0.4}Mn_{0.6}(Fe/Ni)_{0.4}O_{3+\delta}$. Where *a, b, c*, $R_b$ and $R_f$ are the lattice parameters, Bragg factor and fit factors, respectively. $T_C$ is the ferromagnetic Curie temperature, $\theta_p$ is the Curie-Weiss temperature, $\mu_{eff}$ is the effective paramagnetic moment, $M_S$ is the approximate saturation value (at 10 K), $H_C$ is the coercive field (at 10 K), $E_a$ is the activation energy, $\rho^{300}$ is the electrical resistivity (at 300 K) and MR(%) is the maximum values of magnetoresistance (at 125K).

| Compositions | $La_{0.6}Bi_{0.4}Mn_{0.6}Fe_{0.4}O_{3.1}$ | $La_{0.6}Bi_{0.4}Mn_{0.6}Ni_{0.4}O_{2.94}$ |
|---|---|---|
| Space group | *Pnma* | *Pnma* |
| *a* (Å) | 5.514(6) | 5.492(3) |
| *b* (Å) | 7.852(3) | 7.865(3) |
| *c* (Å) | 5.486(2) | 5.502(3) |
| Cell volume (Å³) | 237.52(3) | 237.65(6) |
| $R_b$(%) | 4.22 | 4.59 |
| $R_f$(%) | 9.25 | 8.04 |
| $T_C$ (K) | 95 | 200 |
| $\theta_p$ (K) | 34 | 200 |
| $\mu_{eff}$ ($\mu_B$/f.u.) | 3.17 | 2.85 |
| $H_C$ (Tesla) | 0.25 | 0.0085 |
| $\rho^{300}$ ($\Omega$.cm) | 12 | 34 |
| $E_a$ (meV) | 77 | 82 |
| MR(%) | 2.5 | 20 |

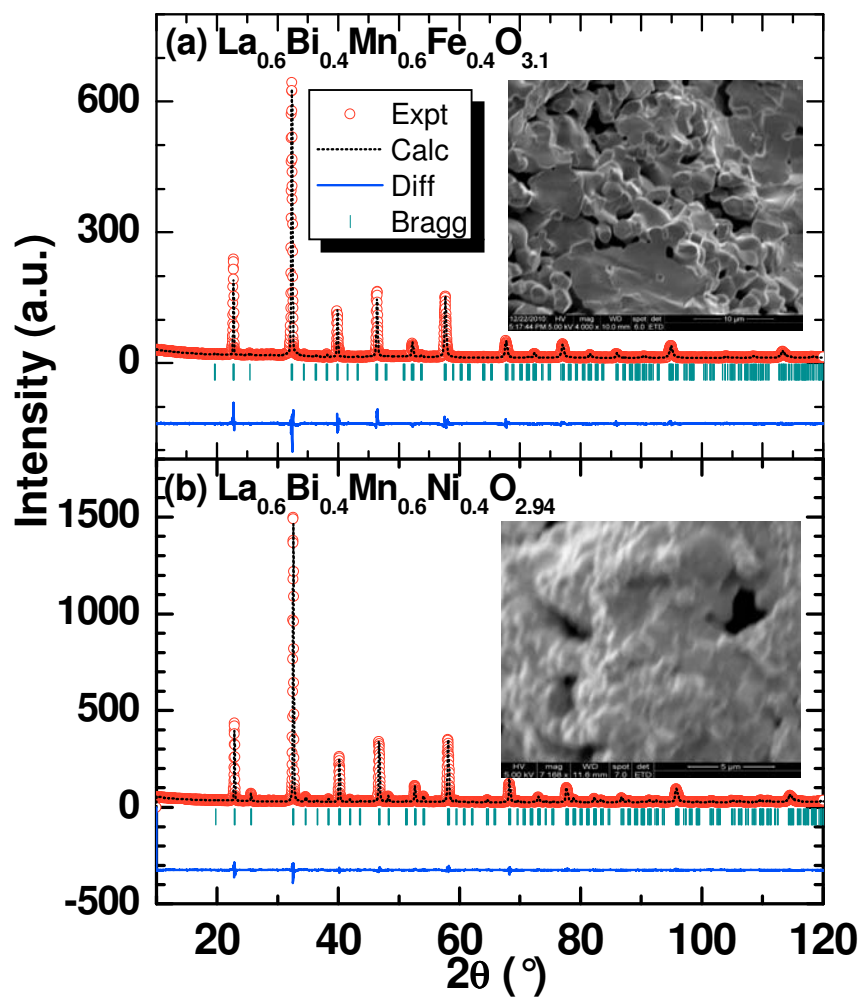

**FIG. 1.** (Color online) Rietveld analysis of XRPD patterns for (a) $La_{0.6}Bi_{0.4}Mn_{0.6}Fe_{0.4}O_{3.1}$ and (b) $La_{0.6}Bi_{0.4}Mn_{0.6}Ni_{0.4}O_{2.94}$ at room temperature. Open symbols are experimental data and the dotted, solid and vertical lines represent the calculated pattern, difference curve and Bragg position respectively. Inset figures show SEM images taken from broken pellets.

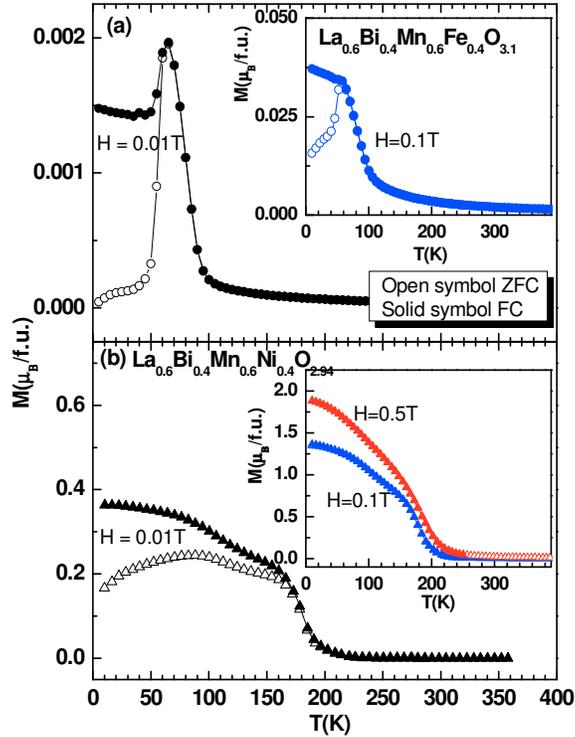

**FIG. 2.** (Color online) Temperature dependent ZFC (open symbol) and FC (solid symbol) magnetization, M(T), for (a) $La_{0.6}Bi_{0.4}Mn_{0.6}Fe_{0.4}O_{3.1}$ and (b) $La_{0.6}Bi_{0.4}Mn_{0.6}Ni_{0.4}O_{2.94}$ in different applied fields (H=0.01, 0.1 and 0.5 Tesla).

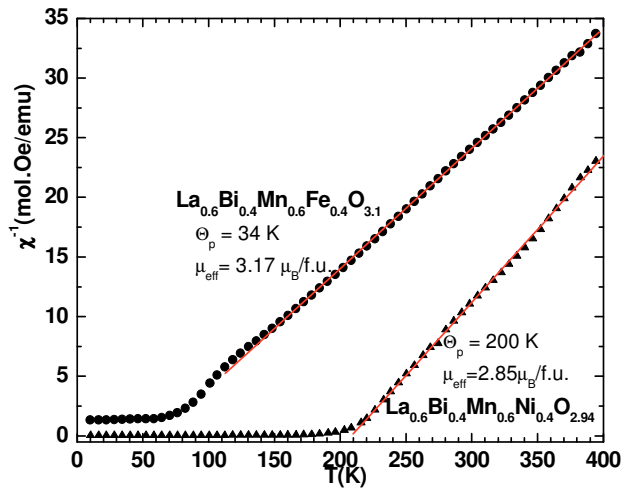

**FIG. 3.** Temperature dependent inverse magnetic susceptibility, $\chi^{-1}$, plot for $La_{0.6}Bi_{0.4}Mn_{0.6}Fe_{0.4}O_{3.1}$ (solid-circle) and $La_{0.6}Bi_{0.4}Mn_{0.6}Ni_{0.4}O_{2.94}$ (solid-triangle). Solid symbols and solid lines represent the experimental data and apparent fit to the Curie-Weiss behaviour.

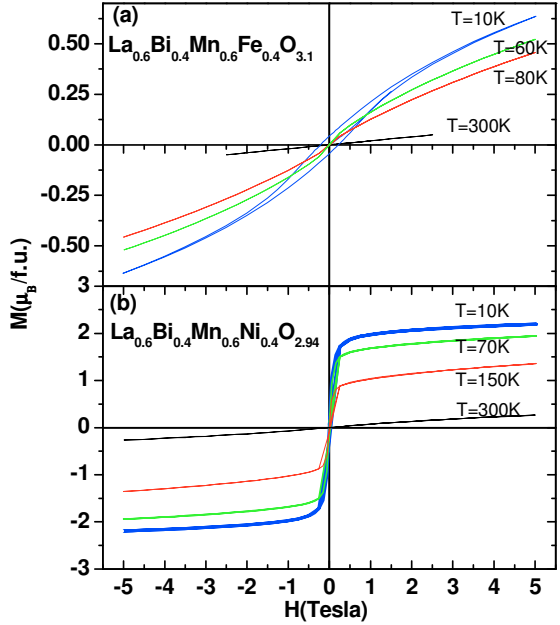

**FIG. 4.** (Color online) Field dependent isothermal magnetic hysteresis, M(H), curves at different temperatures for (a) $La_{0.6}Bi_{0.4}Mn_{0.6}Fe_{0.4}O_{3.1}$ and (b) $La_{0.6}Bi_{0.4}Mn_{0.6}Ni_{0.4}O_{2.94}$

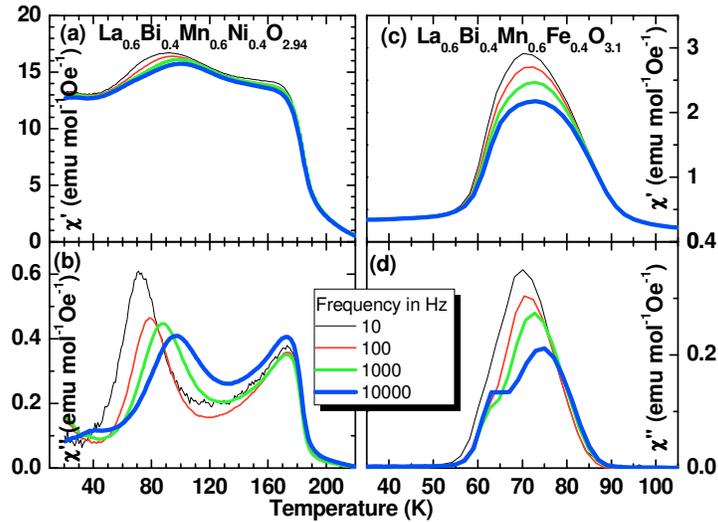

**FIG. 5.** (Color online) Temperature dependent in phase $\chi'$ and out of phase $\chi''$ components of magnetic ac-susceptibility for (a & b) $La_{0.6}Bi_{0.4}Mn_{0.6}Ni_{0.4}O_{2.94}$ and (c & d) $La_{0.6}Bi_{0.4}Mn_{0.6}Fe_{0.4}O_{3.1}$ at different frequencies ($h_{ac}$ = 10 Oe).

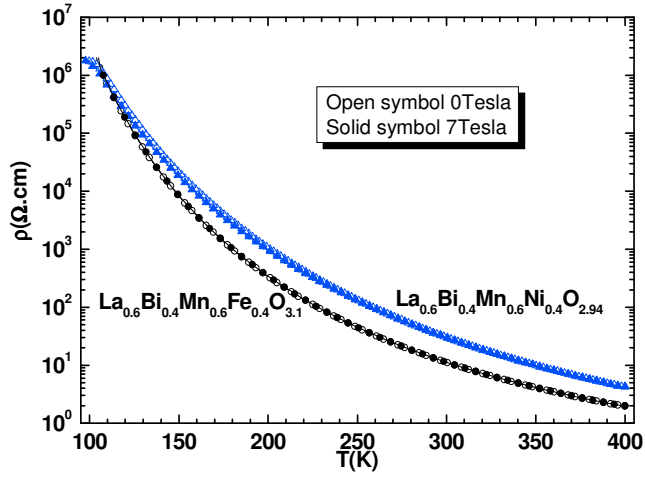

**FIG. 6.** (Color online) Temperature dependent electrical resistivity, ρ(T), for $La_{0.6}Bi_{0.4}Mn_{0.6}Fe_{0.4}O_{3.1}$ (circle) and $La_{0.6}Bi_{0.4}Mn_{0.6}Ni_{0.4}O_{2.94}$ (triangle) in the presence (solid symbol) and absence (open symbol) of magnetic field (7 Tesla).

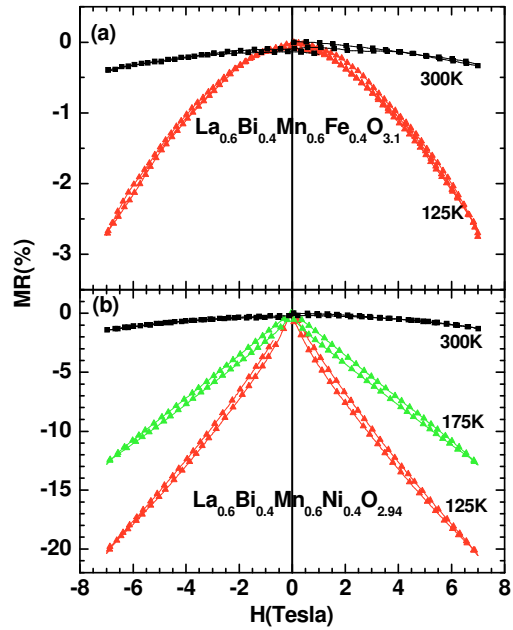

**FIG. 7.** (Color online) The isothermal magnetoresistance MR(%) at different temperatures for (a) $La_{0.6}Bi_{0.4}Mn_{0.6}Fe_{0.4}O_{3.1}$ and (b) $La_{0.6}Bi_{0.4}Mn_{0.6}Ni_{0.4}O_{2.94}$

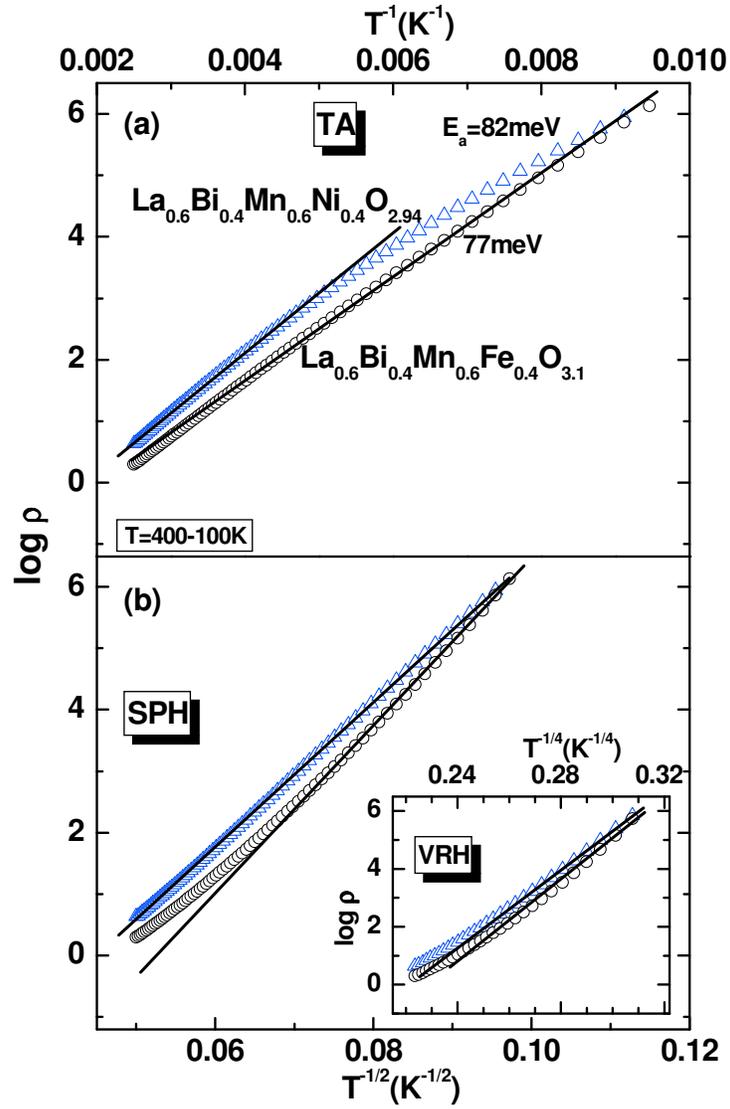

**FIG. 8.** (Color online) Resistivity, $\rho(H=0\ \text{Tesla})$, fit to the relation $\log \rho \propto T^{-1/n}$; where (a) n=1, thermal activation (TA) and (b) n=2, small polaron (SPH) and in the inset n=4, variable range (VRH) hopping model for $La_{0.6}Bi_{0.4}Mn_{0.6}Fe_{0.4}O_{3.1}$ (circle) and $La_{0.6}Bi_{0.4}Mn_{0.6}Ni_{0.4}O_{2.94}$ (triangle).

5